# A New UWB System Based on a Frequency Domain Transformation Of The Received Signal

Karima Ben Hamida El Abri and Ammar Bouallegue

Syscoms Laboratory, National Engineering School of Tunis, Tunisia
Emails: enitkarima@yahoo.fr, ammar.bouallegue@enit.rnu.tn

**Abstract**

*Differential system for ultra wide band (UWB) transmission is a very attractive solution from a practical point of view. In this paper, we present a new direct sequence (DS) UWB system based on the conversion of the received signal from time domain to frequency domain that's why we called FDR receiver. Simulation results show that the prposed receiver structure outperforms the classical differential one for both low and high data rate systems.*

## 1 Introduction

UWB technology has been proposed as a viable solution for high speed indoor short range wireless communications systems due to multipath channel. Moreover, UWB technology can offer simultaneously high data rate and low power implementation. In order to meet the spectrum mask released by the Federal Communications Commission (FCC) and to obtain the adequate signal energy for reliable detection, each information symbol is represented by a train of very short pulses, called monocycles. Each one is located in its own frame.

Regarding demodulation, two types of detectors are considered for UWB systems: coherent and non coherent receiver. The former needs channel information estimation, which is not always achievable, especially at low SNR, even if all the transmitted energy is used for that [9]. In these cases, non-coherent detection is more suitable. That is why, much researches on UWB has focused on non-coherent systems such as: Transmitted Reference (TR) and Differential system. In TR UWB system [10], the transmitted signal consists of a train of pulses pairs. Over each frame, the first pulse is modulated by data. The second one is a reference pulse used for signal detection at the receiver. Reception is made by delaying the received signal and correlating it with the original version. The simplicity of this receiver is very attractive. Nevertheless, TR systems waste half of the energy to transmit reference signals. That's why the TR systems are replaced by differential systems, where detection is achieved by correlating the received signal and its replica delayed by a period D (D can be the symbol period [11], the frame period [12] or a function of chip, symbol and frame period [12] [13]).

In this paper, we introduce a new type of receiver that we called FDR receiver. The structure is very simple. It is based on the projection of the received signal in a basis of functions in order to transform the input of the receiver from time domain to frequency domain, followed by multiplication with a spreading code. With a judicious choice of the basis and the code used, we show that we have not only transform the domain of the received signal. But, we have also transform in someway the PAM (Pulse Amplitude Modulation) modulation used in

DOI : 10.5121/ijwmn.2012.4212 175

International Journal of Wireless & Mobile Networks (IJWMN) Vol. 4, No. 2, April 2012

the transmitter to a PPM (Pulse Position Modulation) modulation. In fact, the useful energy is no longer concentrated in the begining of the frame but it depends on the transmitted data. And this behaviour is similar to PPM modulation.

The remainder of the paper is organized as follows. In section 2, we begin by an overview of the UWB channel model. Then, in section 3, we present the conventionnal differential DS-UWB system followed by description of the proposed receiver in section 4. As for simulation results and comparisons, they are presented in section 5. Finally, a conclusion is given in section 6.

## 2 UWB channel model

The basic conditions of UWB systems differ according to applications. It is based on the conventionnal Saleh and Valenzuela (S-V) channel model [15]. We distinguish two kind of propagation environments: outdoor and indoor propagation. The former is dominated by a direct path while the latter is made of a dense multipath. In this work, we consider the IEEE 802.15 UWB indoor channel [8], where multipath arrivals are grouped into two categories: cluster arrivals, and ray arrivals within each cluster.

$$h(t) = \sum_{l=0}^{L-1} \sum_{k=0}^{K-1} \alpha_{k,l} \delta(t - T_l - \tau_{k,l}) \qquad (1)$$

where:
- $\alpha_{k,l}$ denotes the multipath gain coefficient.
- $T_l$ is the $l^{th}$ cluster arrival time.
- $\tau_{k,l}$ represents the delay of $k^{th}$ multipath component inside the cluster $l$.
- $\delta(t)$ is the Dirac delta function.

The UWB channel given in $(eq.)$ can be modeled as a tapped delay line defined as follows:

$$h(t) = \sum_{l=0}^{L-1} \alpha_l \delta(t - \tau_l) \qquad (2)$$

with:
- $\alpha_l$ denotes attenuation of each path.
- $\tau_l$ represents the delay of $k^{th}$ path. It satisfies $\tau_0 < \tau_1 < \cdots < \tau_L$.

## 3 Related work: Differential DS-UWB system

### 3.1 Modulation

In the UWB transmission, every symbol is transmitted by employing $N_f$ short pulses $\omega_T(t)$, each with an ultra short duration $T_\omega$ of the order of nanosecond and normalized energy. The pulses are transmitted once per frame.

We propose to use a DS-UWB system which is based on a train of short pulses multiplied by a spreading sequence [8]. Each pulse is located in the begining of the frame.
The transmitted signal is given by:

$$s(t) = \sum_j d_j C_{\left\lfloor \frac{j}{N_f} \right\rfloor} \omega_T(t - jT_f) \qquad (3)$$

where:

- $d_j$ is the differentially encoded bit given by: $d_j = d_{j-1} \cdot b_{\left\lfloor \frac{j}{N_f} \right\rfloor}$, where $b_{\left\lfloor \frac{j}{N_f} \right\rfloor}$ represents the random binary data symbol sequence in frame j taking values $\pm 1$, with equal probability.
- $\{C\}$ is the spreading sequence of length $N_f$. Each $C_{\left\lfloor \frac{j}{N_f} \right\rfloor}$ is used to code the pulse contained in the $j^{th}$ frame and it takes values $\pm 1$.
- $T_f$ is the frame duration verifying $T_f \gg T_\omega$.

After propagation in the channel, the received signal $r(t)$ is given by:

$$\begin{aligned} r(t) &= s(t) \otimes h(t) + n(t) \\ &= \sum_j d_j C_{\left\lfloor \frac{j}{N_f} \right\rfloor} \omega_R(t - jT_f) + n(t) \end{aligned} \qquad (4)$$

Where:

- $n(t)$, is an additive white gaussian noise (AWGN) with two side spectral density $\frac{N_0}{2}$ and zero mean.
- $\omega_R(t)$ represents the received waveform of each symbol defined as: $\omega_R(t) = \sum_{l=0}^{L-1} \alpha_l \omega_T(t - \tau_l)$

## 3.2 Demodulation

To detect the emitted symbols $b_m$, we suggest to use a differential receiver based on the correlation of the received signal given in eq.4 with its replica delayed by a frame duration $T_f$.

A block diagram of the receiver is presented in Fig.1 .

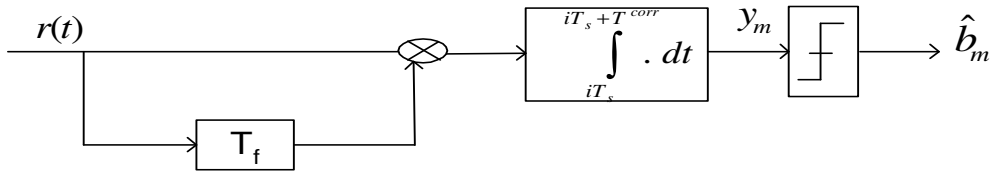

Figure 1: Differential Receiver structure.

The output $y_m$ of the frame differential receiver for the $m^{th}$ bit is given by:

$$\begin{aligned} y_m &= \int_{mT_s}^{(m+1)T_s} r(t).r(t-T_f)dt \\ &= \sum_{j=0}^{N_f-1} \int_{jT_f}^{jT_f+T_{corr}} r(t).r(t-T_f)dt \end{aligned} \quad (5)$$

where:

• $T_{corr}$: is the integration window, $T_{corr} = T_\omega + T_{mds}$ ($T_{mds}$ is the maximum delay spread of the channel).

Let's pose:

$$\begin{aligned} y_j &= \int_{jT_f}^{jT_f+T_{corr}} r(t).r(t-T_f)dt \\ &= s(j) + \sum_{i=1}^{3} n_i(j) \end{aligned} \quad (6)$$

In eq. 6, $s(j)$ is the desired signal in frame j and $\{n_i(j)\}_{i=1:3}$ are noise terms due to signal-noise correlation and noise cross noise correlation.

The expression of the desired signal s(j) is given by:

$$s(j) = b_{\left\lfloor \frac{j}{N_f} \right\rfloor} c_{\left\lfloor \frac{j}{N_f} \right\rfloor} c_{\left\lfloor \frac{j-1}{N_f} \right\rfloor} R_\omega(0) \quad (7)$$

Where $R_\omega(0) = \int_0^{T_{corr}} \omega_T^2(t)dt$.

As for noise terms, they are listed below:

• $n_j(1) = d_j c_{\left\lfloor \frac{j}{N_f} \right\rfloor} \int_0^{T_{corr}} \omega_T(t) n(t+(j-1)T_f)dt$

• $n_j(2) = d_{j-1} c_{\left\lfloor \frac{j-1}{N_f} \right\rfloor} \int_0^{T_{corr}} \omega_T(t) n(t+jT_f)dt$

• $n_j(3) = \int_{jT_f}^{jT_f+T_{corr}} n(t)n(t-T_f)dt$

To find the estimation of the $m^{th}$ bit, we just have to multiply $y_j$ before summation by $c_{\left\lfloor \frac{j}{N_f} \right\rfloor} c_{\left\lfloor \frac{j-1}{N_f} \right\rfloor}$. Then, we only have to take the sign of $y_m \Rightarrow \hat{b}_m = sign(y_m)$

## 4 The proposed system based on a FDR receiver

### 4.1 Modulation

The transmitter structure is the same as the structure described in section 3.
We can write the expression of the emitted signal s(t) as follows:

$$s(t) = \sum_j d_j \omega_T(t - jT_f) \quad (8)$$

Where j is the index of the $j^{th}$ frame, $j \in [0, N * N_f)$ and $N$ represents the total number of emitted bits.

Let's pose: $d_j = c_{\left\lfloor \frac{j}{N_f} \right\rfloor} . b_{\left\lfloor \frac{j}{N_f} \right\rfloor} . d_{j-1}$

For simplicity, we pose: $m = \left\lfloor \frac{j}{N_f} \right\rfloor$ which represents the $m^{th}$ emitted bit, $m \in [0, N)$ and $i = j - mN_f \in [0, N_f)$

And so, the design of the $j^{th}$ code can be obtained as described in Fig.2 .

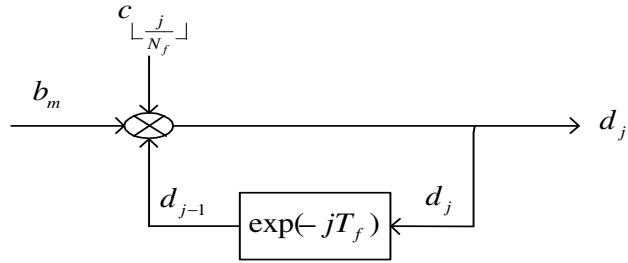

Figure 2: Description of code construction.

By induction, we obtain :

$$d_j = c'_i . b_m^{i+1} . d_{mN_f - 1}$$

where $c'_i = [c'_0, c'_1, \ldots, c'_{N_f - 1}] = \prod_{l=0}^{i} c_l$

Now, we will take the upper value of i to express $d_j$ as a function of $d_{-1}$. In this case ($i = N_f - 1$), we obtain:

$$\begin{aligned} d_j &= d_{mN_f + i} \\ &= d_{(m+1)N_f - 1} \\ &= c'_{N_f - 1} . b_m^{N_f} . d_{mN_f - 1} \end{aligned}$$

$\Rightarrow$

$$\begin{aligned} d_{mN_f - 1} &= c'_{N_f - 1} . b_{m-1}^{N_f} . d_{(m-1)N_f - 1} \\ &= (c'_{N_f - 1})^2 . b_{m-1}^{N_f} . b_{m-2}^{N_f} . d_{(m-2)N_f - 1} \end{aligned}$$

By induction, we find the following relation:

$$d_{mN_f - 1} = (c'_{N_f - 1})^m . (\prod_{l=0}^{m-1} b_l)^{N_f} . d_{-1}$$

And so, the expression of $d_j$ becomes:

$$d_j = c'_i . b_m^{i+1} . d_{mN_f-1} = c'_i . b_m^i . a_m$$

Where:

$$a_m = (c'_{N_f-1})^m . (\prod_{l=0}^{m-1} b_l)^{N_f} . d_{-1} . b_m \qquad (9)$$

Hence, the expression of s(t) can be rewritten as follows:

$$\begin{aligned} s(t) &= \sum_j d_j \omega(t - jT_f) \\ &= \sum_j a_m . c'_i . b_m^i . \omega(t - jT_f) \end{aligned} \qquad (10)$$

### 4.2 Demodulation

After propagation in the UWB channel, the transmitted signal arrives at the receiver in distorted waveforms and so, the received signal r(t) is given by:

$$\begin{aligned} r(t) &= s(t) \otimes h(t) + n(t) \\ &= \sum_j a_m \, c'_i \, b_m^i \omega_R(t - jT_f) + n(t) \end{aligned} \qquad (11)$$

Where:
- $\otimes$ represents the convolution operator.
- $h(t)$ is the channel impulse response.
- $n(t)$ is an Additive White Gaussian Noise, with power density $\frac{N_0}{2}$.
- $\omega_R(t)$ is the received waveform after propagation in the channel,
$$\omega_R(t) = \omega_T(t) \otimes h(t).$$
- m and i are defined for simplification. They are given by: $m = \left\lfloor \frac{j}{N_f} \right\rfloor$ and $i = j - mN_f$

The receiver model is based on the projection of the received signal r(t) in the frequency domain. The receiver structure is described in Fig.3 .

At the input of the receiver, the signal $r_m(t)$ which represents the received signal in the $m^{th}$ bit, is composed of $N_f$ signal $r_m(l,t)$ which corresponds to the signal in the frame l.

$$\begin{aligned} r_m(t) &= \sum_{l=0}^{N_f-1} r_m(l,t) \\ &= \sum_{l=0}^{N_f-1} r_m(t + l.T_f) \end{aligned} \qquad (12)$$

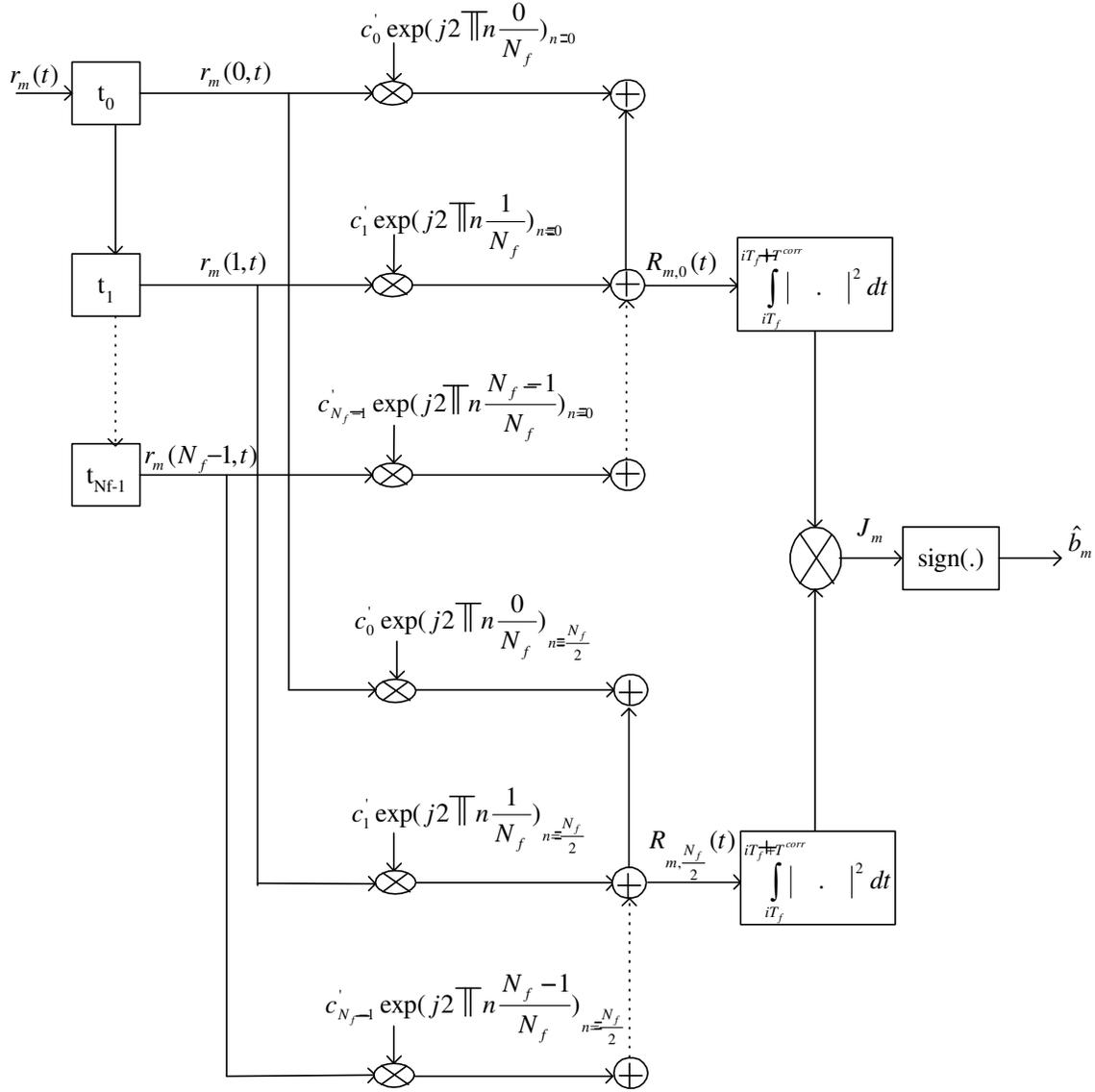

Figure 3: The FDR receiver structure.

Each portion of $r_m(t)$ can be written as:
$$r_m(l, t) = s_m(l, t) + n_m(l, t)$$

Where $s_m(l, t)$ is the desired signal in the $m^{th}$ bit, in the frame j. It is given by:

$$\begin{aligned} s_m(l, t) &= s_m(t + l.T_f) \\ &= a_m . c_i' . b_m^i . \omega_R(t) \end{aligned}$$

Then, we project each portion of $r_m(t)$ in the new basis and multiply it by the code $c_l'$.

Hence, we obtain the following signal for bit m:

$$R_{m,n}(t) = \sum_{l=0}^{N_f-1} c'_l\, r(l,t)\, exp(j2\pi \frac{nl}{N_f})$$
$$= S_{m,n}(t) + N_{m,n}(t) \quad (13)$$

Where:
- $n \in [0, N_f)$
- $S_{m,n}(t)$ and $N_{m,n}(t)$ are respective projections of $s(l,t)$ and $n(l,t)$ in the new basis.

We show in the Appendix that the estimation of bit m ($\hat{b}_m$) is specified using only two values of $R_{m,n}(t)$ which are: $R_{m,0}(t)$ and $R_{m,\frac{N_f}{2}}(t)$. In fact, the useful energy is concentrated in $R_{m,0}(t)$ if the transmitted bit $b_m = 1$ or in $R_{m,\frac{N_f}{2}}(t)$ if the transmitted bit $b_m = -1$. As for noise component, they are located in the other cases. We have:

- $b_m = 1 \Rightarrow$

$$|R_{m,n}(t)|^2 = N_f^2\, \omega_R^2(t) + a_m N_f \omega_R(t)(N_{m,n}(t) + N_{m,n}(t)^*) + |N_{m,n}(t)|^2, if\ n = 0$$
$$|R_{m,n}(t)|^2 = |N_{m,n}(t)|^2, if\ n \neq 0$$

- $b_m = -1 \Rightarrow$

$$|R_{m,n}(t)|^2 = N_f^2\, \omega_R^2(t) + a_m N_f \omega_R(t)(N_{m,n}(t) + N_{m,n}(t)^*) + |N_{m,n}(t)|^2, if\ n = \frac{N_f}{2}$$
$$|R_{m,n}(t)|^2 = |N_{m,n}(t)|^2, if\ n \neq \frac{N_f}{2}$$

This behaviour is similar to a PPM modulation using frames of duration $N_f T_f$ and pulses having energy $N_f$ times more than the transmitted pulses. And so, the detection is based on an energy collector intergator.
Hence, we transform a PAM modulation to a PPM modulation. This explain the performance of the proposed system using the FDR receiver which outperforms the differential receiver.
Therefore, to decide of the value of $\hat{b}_m$, we have to use the following criterion:

$$J_m = \int_0^{T^{corr}} |R_{m,0}(t)|^2 dt - \int_0^{T^{corr}} \left|R_{m,\frac{N_f}{2}}(t)\right|^2 dt$$
$$= \int_0^{T^{corr}} \left[|R_{m,0}(t)|^2 - \left|R_{m,\frac{N_f}{2}}(t)\right|^2\right] dt \quad (14)$$

The decision is given by the sign of the criterion $J_m$.

## 5 Simulation

In this section, we present the parameters used in the simulation of the two systems implemented: the DS-UWB system using the differential receiver and the DS-UWB system using the proposed FDR receiver. Results are given in terms of Bit Error Rate (BER) as a function of Signal to Noise Ratio (SNR).

The simulations are performed through numerical Monte Carlo simulations. In each trial, the following suppositions are made:
 • The monocycle $\omega_T(t)$ is normalized so that the total symbol energy is unity, with duration $T_\omega = 0.8 ns$.
 • The spreading codes are generated randomly from $\pm 1$, with equal probability.
 • We simulate the multipath channel using the model CM1 from [8]. The channel is assumed to be time invariant within a burst of symbols. The maximum delay spread of the channel is $5 ns$. To avoid inter frame interference, we truncated the channel to $\frac{T_f}{2}$.

First, we begin by evaluating the proposed frequency domain receiver (FDR) performance for a fixed data rate, with different pulse shape. The result is drawn in Fig. 4.

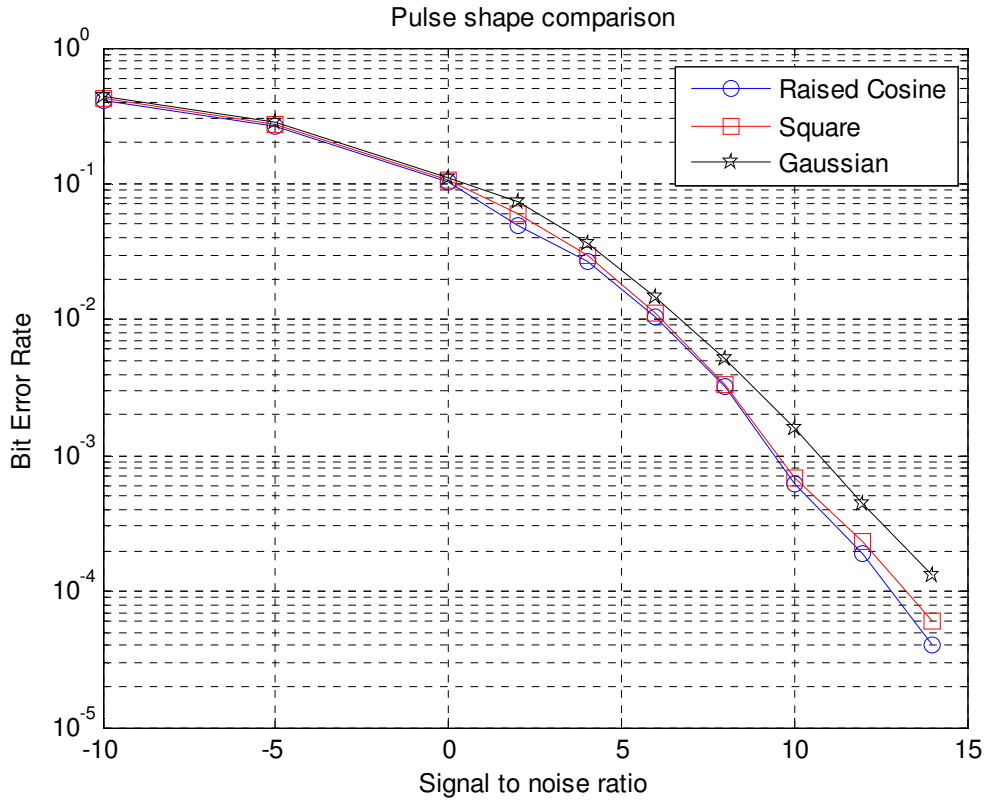

Figure 4: Performance of the FDR receiver for different pulse shape

As shown in Fig. 4, the FDR receiver offers good performance in terms of BER, for the different pulse shapes. We can notice that the square pulse and the raised cosine pulse (with roll off factor $\alpha$=0.6) offer better performance than the gaussian pulse. In fact, we get an improvment of about $1dB$ when we use the raised cosine pulse. Furthermore, the raised cosine shape is recommended by the FCC because it is the most suitable with the mask imposed by the commission. That's why, we consider the raised cosine pulse as pulse reference in the next simulation.

In wireless communications, there is always a compomise between performance and data rate. That's why, we test the performance of the proposed system for 3 different data rate: $6$, 12 and $25 Mbit/s$. The results are given in Fig. 5.

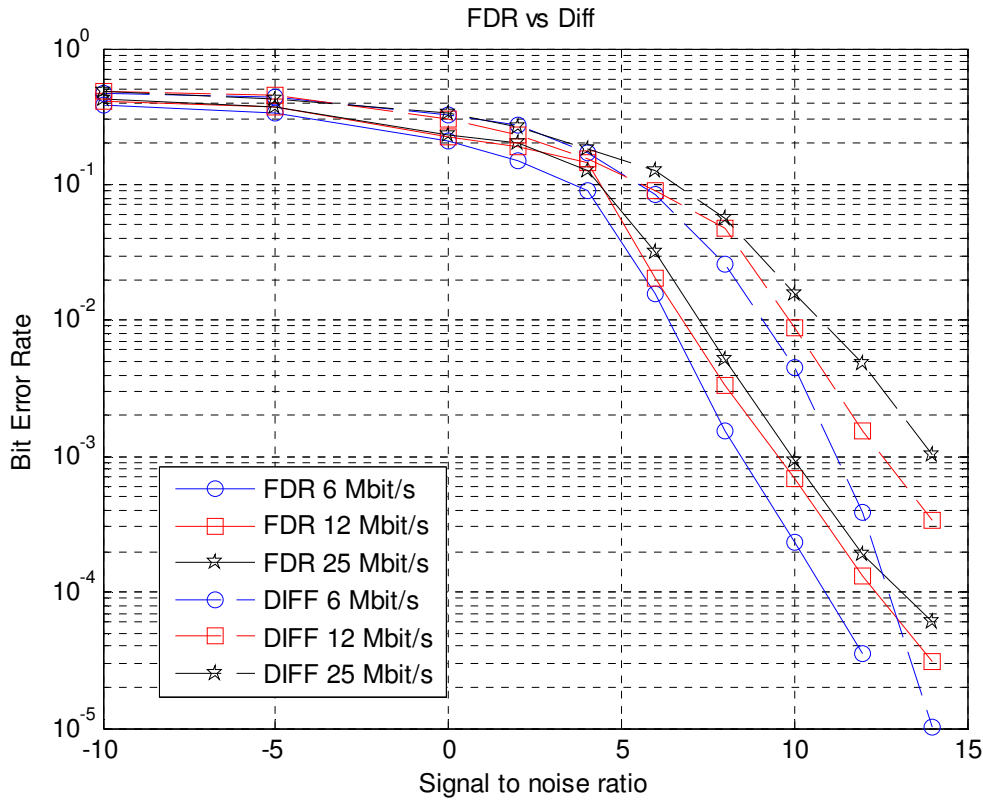

Figure 5: Performance comparision for different data rate

As we can see in Fig. 5 and as expected, the bit error rate increases with the increase of the data rate for both systems: the DS-UWB system using the FDR receiver and the system using the classical differential receiver. But, we can note that both of them are robust to high data rate. On the other hand, the FDR receiver outperforms the differential receiver. For example, for $6 Mbit/s$, and to achieve a BER= $10^{-3}$, the FDR receiver needs $8 dB$ while the differential reciever needs $11 dB$, and so we get a gain of about $3 dB$. We can also notice for example, a loss of about $2 dB$ when increasing the data rate from $6$ to $25 Mbps$, for the FDR receiver. As for the differential receiver we can perceive a loss of $3 dB$.

Thereby, the FDR receiver is more suitable for high data rate.

## 6 Conclusion

In this paper, a new receiver structure called $FDR$ is proposed for DS-UWB system. It is based on the transformation of the signal arriving at the receiver input from time domain to frequency domain.

We show through simulations that the proposed receiver gives good performance in terms of BER. Besides, the proposed receiver outperforms the differential receiver for the different data rate tested. We have an improvement of about $3dB$ compared to the differential scheme.

As a future work, we propose to evaluate the robustness of the FDR receiver to inter frame interference and multi user interference.

## Appendix

The expression of $S_{m,n}(t)$ can be written as:

$$S_{m,n}(t) = \sum_{l=0}^{N_f-1} c'_l s_m(l,t) \exp(j2\pi \frac{nl}{N_f})$$

$$= \sum_{l=0}^{N_f-1} c'_l a_m c'_i b^i_m \omega_R(t) \exp(j2\pi \frac{ln}{N_f})$$

$$= \sum_{l=0}^{N_f-1} a_m b^i_m \omega_R(t) \exp(j2\pi \frac{nl}{N_f}), (c'_l = c'_i \text{ car } i = l - mN_f)$$

$$= a_m \omega_R(t) \sum_{l=0}^{N_f-1} b^i_m \exp(j2\pi \frac{nl}{N_f})$$

We can write $b^i_m$ as an exponential as follows:

$$b^i_m = \exp(j2\pi i \frac{b_m-1}{4})$$

In fact, if $b_m = 1$, we obtain $j2\pi i \frac{b_m-1}{4} = 0$ and so $b_m = \exp(0) = 1$

As for the case where $b_m = -1$, we get $j2\pi i \frac{b_m-1}{4} = \exp(-j\pi)$ and so $b_m = -1$

Using this tranforamtion of the expression of $b^i_m$, $S_{m,n}(t)$ becomes:

$$S_{m,n}(t) = a_m . \omega_R(t) \sum_{l=0}^{N_f-1} \exp\left[j2\pi \left(\frac{nl}{N_f} + i\frac{b_m-1}{4}\right)\right]$$

From the last expression, we can notice that the value of $S_{m,n}(t)$ depends on the exponential term. That's why, we will try to find the value of this term below. Two cases have to be evaluated: $b_m = 1$ and $b_m = -1$.

**- First case: $b_m = 1$**

In this case, $S_{m,n}(t)$ is given as follows:

$$S_{m,n}(t) = a_m \omega_R(t) \sum_{l=0}^{N_f-1} \exp\left[j2\pi \frac{nl}{N_f}\right]$$

- If $n = 0$ then $S_{m,0}(t) = a_m N_f \omega_R(t)$

- If $n \neq 0$ then $S_{m,n}(t) = a_m \, \omega_R(t) \dfrac{1-\exp\left[j2n\pi\frac{lN_f}{N_f}\right]}{1-\exp\left[j2\pi\frac{nl}{N_f}\right]} \Rightarrow S_n(t) = 0$

**- Second case: $b_m = -1$**
In this case, $S_{m,n}(t)$ can be written as:
$$S_{m,n}(t) = a_m \, \omega_R(t) \sum_{l=0}^{N_f-1} \exp\left[j2\pi\left(\frac{nl}{N_f} - \frac{i}{2}\right)\right]$$

- If $j2\pi\left(\frac{nl}{N_f} - \frac{i}{2}\right) = 0 \Rightarrow \frac{nl}{N_f} - \frac{i}{2} = 0 \Rightarrow n = \frac{l-mN_f}{2l} N_f \Rightarrow n = \frac{N_f}{2} - \frac{m}{2l}. N_f^2 \in [0, N_f) \Rightarrow n = \frac{N_f}{2}$

Hence, if $n = \frac{N_f}{2}$, then $S_{m,\frac{N_f}{2}}(t) = a_m \, N_f \, \omega_R(t)$

- If $n \neq \frac{N_f}{2} \Rightarrow$
$$S_{m,n}(t) = a_m \, \omega_R(t) \sum_{l=0}^{N_f-1} \exp\left[j2\pi\left(\frac{nl}{N_f} - \frac{i}{2}\right)\right]$$
$$= a_m \, \omega_R(t) \dfrac{1-\exp\left[j2\pi N_f\left(\frac{nl}{N_f} - \frac{i}{2}\right)\right]}{1-\exp\left[j2\pi\left(\frac{nl}{N_f} - \frac{i}{2}\right)\right]}$$

We know that:
$$0 \leq n < N_f$$
$$0 \leq \frac{n}{N_f} < 1$$
$$\Rightarrow \quad 0 \leq \frac{nl}{N_f} < l < N_f$$
$$(0 < i < N_f)$$
$$\Rightarrow \quad 0 \leq \frac{nl}{N_f} - \frac{i}{2} < N_f - \frac{i}{2}, (i \in [0, N_f))$$
$$\Rightarrow \quad \exp\left[j2\pi N_f\left(\frac{nl}{N_f} - \frac{i}{2}\right)\right] = 1$$

Hence, $S_{m,n}(t) = 0$ if $n \neq \frac{N_f}{2}$

If we resume what was said above, $S_{m,n}(t)$ depends on both $b_m$ and $n$. It is given by expressions below.

- $b_m = 1 \Rightarrow$
$$S_{m,0}(t) = a_m \, N_f \, \omega_R(t)$$
$$S_{m,n}(t) = 0 \; if \; n \neq 0$$

- $b_m = -1 \Rightarrow$

$$S_{m,\frac{N_f}{2}}(t) = a_m N_f \omega_R(t)$$

$$S_{m,n}(t) = 0 \ if \ n \neq \frac{N_f}{2}$$

Therefore, the decision can be given by $S_{m,n}(t)$ and so $R_{m,n}(t)$. It is a soft decision because $\hat{b}_m$ is determinated by two values of $R_{m,n}(t)$: $n = \frac{N_f}{2}$ and $n = 0$.

First, we suppose that $N_f$ is even.

Since $R_{m,n}(t)$ is complex, we calculate its norm to give the criterion adopted for decision.

$$\begin{aligned}|R_{m,n}(t)|^2 &= (S_{m,n}(t) + N_{m,n}(t))(S_{m,n}(t) + N_{m,n}(t))^* \\ &= |S_{m,n}(t)|^2 + S_{m,n}(t)^* N_{m,n}(t) + S_{m,n}(t) N_{m,n}(t)^* + |N_{m,n}(t)|^2\end{aligned}$$

- If $b_m = 1$ then:

$$|R_{m,n}(t)|^2 = N_f^2 \omega_R^2(t) + a_m N_f \omega_R(t)(N_{m,n}(t) + N_{m,n}(t)^*) + |N_{m,n}(t)|^2, if \ n = 0$$

$$|R_{m,n}(t)|^2 = |N_{m,n}(t)|^2, if \ n \neq 0$$

- If $b_m = -1$ then:

$$|R_{m,n}(t)|^2 = N_f^2 \omega_R^2(t) + a_m N_f \omega_R(t)(N_{m,n}(t) + N_{m,n}(t)^*) + |N_{m,n}(t)|^2, if \ n = \frac{N_f}{2}$$

$$|R_{m,n}(t)|^2 = |N_{m,n}(t)|^2, if \ n \neq \frac{N_f}{2}$$

We note that the useful energy is concentrated whether in $R_{m,0}(t)$ if $b_m = 1$, or in $R_{m,\frac{N_f}{2}}(t)$ if $b_m = -1$. As for noise components, they are located in the other cases.

This behaviour is similar to a PPM modulation using frames of duration $N_f T_f$ and pulses having energy $N_f$ times more than the transmitted pulses. And so, the detection is based on an energy collector intergator.

Therefore, to decide of the value of $\hat{b}_m$, we have to use the following criterion:

$$\int_0^{T^{corr}} |R_{m,0}(t)|^2 dt \geq \int_0^{T^{corr}} \left|R_{m,\frac{N_f}{2}}(t)\right|^2 dt \ \Rightarrow \ \hat{b}_m = 1$$

$$\int_0^{T^{corr}} |R_{m,0}(t)|^2 dt \leq \int_0^{T^{corr}} \left|R_{m,\frac{N_f}{2}}(t)\right|^2 dt \ \Rightarrow \ \hat{b}_m = -1$$